\documentclass{aa}
\usepackage{graphicx}
\usepackage{txfonts}
\usepackage{natbib}

\begin{document}


\def\be{\begin{equation}}
\def\ee{\end{equation}}
\def\bea{\begin{eqnarray}}
\def\eea{\end{eqnarray}}
\def\c{\cite}
\def\nn{\nonumber}
\def\mcr{{{\rm M_{cr}}}}
\def\xo{{X_{o}}}
\def\dm{\Delta {\rm M}}
\def\ms{{\rm M_{\odot}}}
\def\mb{m_{B}}
\def\bo{B_{o}}
\def\cxo{C(x_{o})}
\def\rsix{R_{6}}
\def\vr{v_{{\rm r}}}
\def\vro{v_{{\rm ro}}}
\def\vt{v_{\theta}}
\def\po{\ifmmode P_{o} \else $P_{o}$ \fi}

\def\et{ {\it et al.}}
\def\la{ \langle}
\def\ra{ \rangle}
\def\ov{ \over}
\def\ep{ \epsilon}

\def\th{\theta}
\def\ga{\gamma}
\def\Ga{\Gamma}
\def\la{\lambda}
\def\si{\sigma}
\def\al{\alpha}
\def\pa{\partial}
\def\de{\delta}
\def\De{\Delta}
\def\rsr{{r_{s}\over r}}
\def\rmo{{\rm R_{M0}}}
\def\rrm{{R_{{\rm M}}}}
\def\rra{{R_{{\rm A}}}}

\def\mdot{\ifmmode \dot M \else $\dot M$\fi}    
\def\mxd{\ifmmode \dot {M}_{x} \else $\dot {M}_{x}$\fi}
\def\med{\ifmmode \dot {M}_{Edd} \else $\dot {M}_{Edd}$\fi}
\def\bff{\ifmmode B_{{\rm f}} \else $B_{{\rm f}}$\fi}

\def\apj{\ifmmode ApJ \else ApJ \fi}    
\def\apjl{\ifmmode  ApJ \else ApJ \fi}    %
\def\aap{\ifmmode A\&A \else A\&A\fi}    %
\def\mnras{\ifmmode MNRAS \else MNRAS \fi}    %
\def\nat{\ifmmode Nature \else Nature \fi}
\def\prl{\ifmmode Phys. Rev. Lett. \else Phys. Rev. Lett.\fi}
\def\prd{\ifmmode Phys. Rev. D. \else Phys. Rev. D.\fi}

\def\ms{\ifmmode M_{\odot} \else $M_{\odot}$\fi}    
\def\na{\ifmmode \nu_{A} \else $\nu_{A}$\fi}    
\def\nk{\ifmmode \nu_{K} \else $\nu_{K}$\fi}    
\def\ns{\ifmmode \nu_{{\rm s}} \else $\nu_{{\rm s}}$\fi}
\def\no{\ifmmode \nu_{1} \else $\nu_{1}$\fi}    
\def\nt{\ifmmode \nu_{2} \else $\nu_{2}$\fi}    
\def\ntk{\ifmmode \nu_{2k} \else $\nu_{2k}$\fi}    
\def\dnmax{\ifmmode \Delta \nu_{max} \else $\Delta \nu_{2max}$\fi}
\def\ntmax{\ifmmode \nu_{2max} \else $\nu_{2max}$\fi}    
\def\nomax{\ifmmode \nu_{1max} \else $\nu_{1max}$\fi}    
\def\nh{\ifmmode \nu_{\rm HBO} \else $\nu_{\rm HBO}$\fi}    
\def\nqpo{\ifmmode \nu_{QPO} \else $\nu_{QPO}$\fi}    
\def\nz{\ifmmode \nu_{o} \else $\nu_{o}$\fi}    
\def\nht{\ifmmode \nu_{H2} \else $\nu_{H2}$\fi}    
\def\ns{\ifmmode \nu_{s} \else $\nu_{s}$\fi}    
\def\nb{\ifmmode \nu_{{\rm burst}} \else $\nu_{{\rm burst}}$\fi}
\def\nkm{\ifmmode \nu_{km} \else $\nu_{km}$\fi}    
\def\ka{\ifmmode \kappa \else \kappa\fi}    
\def\dn{\ifmmode \Delta\nu \else \Delta\nu\fi}    

\def\rs{\ifmmode {R_{s}} \else $R_{s}$\fi}    
\def\ra{\ifmmode R_{A} \else $R_{A}$\fi}    
\def\rso{\ifmmode R_{S1} \else $R_{S1}$\fi}    
\def\rst{\ifmmode R_{S2} \else $R_{S2}$\fi}    
\def\rmm{\ifmmode R_{M} \else $R_{M}$\fi}    
\def\rco{\ifmmode R_{co} \else $R_{co}$\fi}    
\def\ris{\ifmmode {R}_{{\rm ISCO}} \else $ {\rm R}_{{\rm ISCO}} $\fi}
\def\rsix{\ifmmode {R_{6}} \else $R_{6}$\fi}
\def\rinfty{\ifmmode {R_{\infty}} \else $R_{\infty}$\fi}
\def\rinfsix{\ifmmode {R_{\infty6}} \else $R_{\infty6}$\fi}

\def\rxj{\ifmmode {RX J1856.5-3754} \else RX J1856.5-3754\fi}
\def\1739{\ifmmode {XTE  J1739-285} \else XTE  J1739-285\fi}
 \def\exo{\ifmmode {EXO 0748-676} \else EXO 0748-676\fi}


\title{A model for upper kHz QPO coherence of accreting neutron star}

\author{J. Wang$^1$, C. M. Zhang$^1$, Y. H. Zhao$^1$,
Y. F. Lin,$^2$  H. X. Yin$^3$, L. M. Song$^4$\\}

\institute{ $^{1}$ National Astronomical Observatories,  Chinese
Academy of
Sciences, Beijing 100012, P. R. China, jwang@bao.ac.cn, zhangcm@bao.ac.cn \\
$^2$Department of Physics and Tsinghua Center for Astrophysics,
Tsinghua University, Beijing 100084, China\\
$^{3}$ School of Space Science and Physics, Shandong University,
Weihai 264209, China\\
$^{4}$ Institute of High Energy Physics, Chinese Academy of
Sciences, Beijing 100049,   China}


\abstract {We investigate the coherence of the twin kilohertz
quasi-periodic oscillations (kHz QPOs) in the low-mass X-ray binary
(LMXB) theoretically. The profile of upper kHz QPO, interpreted as
Keplerian frequency, is ascribed to the radial extent of the kHz QPO
emission region, associated with the transitional layer at the
magnetosphere-disk boundary, which corresponds to the coherence of
upper kHz QPO. The theoretical model for Q-factor of upper kHz QPO
is applied to the observational data of five Atoll and five Z
sources, and the consistence is implied.
 }

\maketitle
\section{Introduction}

The launch of the Rossi X-ray Timing Explorer (RXTE) led to the
discovery of Kilohertz quasi-periodic oscillations (kHz QPOs) of
X-ray spectra in low-mass X-ray binaries (LMXBs), i.e., narrow
features in their power density spectra (PDS) (van der Klis 2000,
2006). These frequencies, in the range of $200 \sim 1300$ Hz, are
the same order of the dynamical time-scales of the innermost regions
of the accretion flow around the stellar mass compact objects (van
der Klis 2006, 2008), which may carry imprints of strong field
general relativity phenomena {\bf (e.g. Kluzniak, Michelson, Wagoner
1990; Miller, Lamb \& Psaltis 1998; Kluzniak 1998; Abramowicz et al.
2003)}. Owing to the expected links with the orbital motion, most
works about the kHz QPOs focus on the explanation for the emission
of these frequencies of the orbital Keplerian motion (e.g. Miller,
Lamb \& Psaltis 1998; Stella \& Vietri 1998, 1999; Kluzniak \&
Abramowicz 2001; Abramowicz et al. 2003; Zhang 2004).
In addition, the kHz QPOs were found to occur in twin peaks usually
(upper $\nu_2$ and lower $\nu_1$ frequency). They behave in a rather
regular way and follow the tight correlations between their
frequencies and other observed characteristic frequencies (see, e.g.
Psaltis et al. 1998, 1999a; Psaltis, Belloni \& van der Klis 1999b;
Stella, Vietri \& Morsink 1999; Belloni, Psaltis \& van der Klis
2002; Titarchuk \& Wood 2002; M$\acute{e}$ndez \& van der Klis 1999,
2000; M$\acute{e}$ndez et al. 2001; {\bf Yu W. F., van der Klis M.,
Jonker P. G. 2001; Yu W. F., van der Klis M. 2002}). Moreover, the
correlation between the upper frequency and lower frequency across
different sources can be roughly fitted by a power-law function (see
e.g. Psaltis et al. 1998, 1999a; Zhang et al. 2006a), and also by a
linear model (see Belloni, Mendez \& Homan 2005, 2007).

\begin{figure}
\includegraphics[width=8cm]{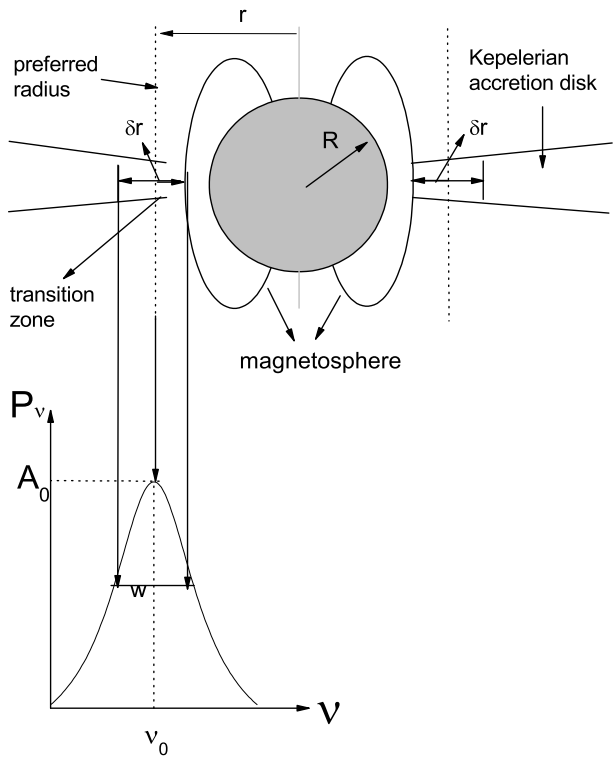}
\caption{The illustration of Lorentzian function as described in Eq.
(1) is associated with the sketch map of the magnetosphere-disk
transition layer. The labels $A_{0}$, $\nu_{0}$ and $w$ are noted in
Lorentzian function Eq.(1). In the upper panel of figure, the
stellar radius R, magnetosphere radius r and magnetosphere-disk
transition layer width $\delta r$ are noted, and the kHz QPO profile
and its corresponding positions in the transition layer are
presented.}\label{model}
\end{figure}

The kHz QPOs in LMXB are typical time variable signals and peaks
with some width in PDS, with the particular case of the sharp
coherent pulsation 401 Hz and a near 401 Hz X-ray burst oscillation
frequency are found in SAX J1808.4-3658 (Chakrabarty et al. 2003;
Wijnands et al. 2003; Wijnands 2005), whose profiles can be
described by a Lorentzian Function (see the lower panel in Fig.
\ref{model}),
\begin{equation}
P_{\nu}\propto A_{0} w/[(\nu-\nu_0)^2+(w/2)^2],
\end{equation}
where $\nu_0$ is the peak frequency, $w$ is the full width at
half-maximum (FWHM), and $A_0$ is the amplitude of this signal. The
ratio of these two quantities are the quality factor, i.e.,
\begin{equation}
Q \equiv \frac{\nu_0}{w}.
\end{equation}
This function is characterized by three characteristic quantities
which are used to describe the profile of these signals, i.e.
centroid frequency (i.e. peak frequency $\nu_0$), quality factor (Q
$\equiv \nu_0~/$ FWHM) and the fractional root-mean-squared (rms).
The quality factor characterizes the coherence of a QPO signal,
while the rms represents a measure of the signal strength, which is
proportional to the square root of the peak power contribution to
the PDS. The large RXTE archive makes possible searches for
systematic correlations between these three characteristic
quantities. Using data from RXTE, Barret et al. (2005a) studied 4U
1608-52 and revealed a positive correlation between lower frequency
and its quality factor, up to a maximum of about Q $\sim$ 200.
Motivated by this idea, Barret, Olive \& Miller (2005b, 2006)
studied, in a systematic way, the QPO properties of 4U 1636-536 and
the dependency of quality factor and rms on frequency. It is shown
that quality factors for the lower and upper kHz QPOs of 4U 1636-536
follow different tracks in a Q versus frequency plot, i.e. quality
factor for the lower kHz QPO increases with frequency up to 850 Hz
(Q $\sim$ 200) and drops precipitously to the highest detected
frequencies $\sim$ 920 Hz (Q $\sim$ 50), while that of the upper kHz
QPO increases steadily all the way to the highest detectable QPO
frequency. Moreover, quality factor of the lower QPO is higher than
that of the upper (Barret, Olive \& Miller 2005b,c; 2006). In
addition, the fractional rms amplitudes of both the upper and lower
kHz QPOs increase then decrease steadily towards higher frequencies,
with a ceiling in lower kHz QPO (Barret, Olive \& Miller 2005b,c;
2006). A similar behavior was seen from 4U 1608-52 (Barret et al.
2005a). The rough similarity also was extended to 4U 1735-44, 4U
1728-34 (Barret, Olive \& Miller 2006; Boutelier, Barret \& Miller
2009; M$\acute{e}$ndez 2006; T$\ddot{o}$r$\ddot{o}$k 2009). The
complete list of references on QPOs from these sources is available
in van der Klis (2006). Then T$\ddot{o}$r$\ddot{o}$k (2009) studied
the difference in rms amplitude between the upper and lower kHz QPOs
as a function of the frequency ratio ($\nu_2/\nu_1$). They found
that the rms amplitudes of the twin peaks become equal when the
frequencies of the oscillations pass through a certain ratio
($\nu_2/\nu_1$), which is roughly the same for each of the sources.
It is also predicted that in a more general context, the behaviour
of the amplitude difference suggests a possible energy interchange
between the upper and lower QPO modes (T$\ddot{o}$r$\ddot{o}$k
2009).

Theoretically, the quality factor and rms amplitude have also been
studied but not yet satisfactorily settled. In the past years,
several works, focussing on the quality factor and rms as separated
functions of frequencies, have been discussed on these properties,
and possible consequences for various QPO models have been outlined
(see, e.g., Barret et al. 2005a; M$\acute{e}$ndez 2006; Barret,
Olive \& Miller 2006, for further information and references).
M$\acute{e}$ndez (2006) considered the relation between the
innermost stable circular orbit (ISCO) and the drop of QPO coherence
and rms. Barret, Olive \& Miller (2006) discussed the implications
of their results and showed how the high-frequency drop-off in
quality factor can be accommodated quantitatively in a toy model
based on the approach to the ISCO. In their toy model, the change of
quality factor was ascribed to three basic parts: (1) the finite
extent in radial direction($\Delta r_{orb}$), (2) the radial drift
($\Delta r_{drift}$) during the lifetime of the oscillation, (3) the
finite time itself (Barret, Olive \& Miller 2006). It was also
proposed that a drop in the amplitude and quality factor of the QPOs
at some limiting frequency is a possible signature of the ISCO
(Miller, Lamb \& Psaltis 1998; Barret, Olive \& Miller 2005b,c).

The abrupt rise of X-ray flux, as a cause of QPO, may be related to
the abrupt transformation of environment in accretion process, which
is ascribed to the interaction of magnetosphere dominated regime and
gravitational induced energy-momentum dominated disk.
In this regards, the interpretation of the twin kHz QPOs has been
ascribed to the orbital Keplerian frequency at magnetosphere-disk
boundary for upper kHz QPO and MHD Alfv$\grave{e}$n wave propagation
frequency there for lower kHz QPO, respectively (Zhang 2004; Zhang
et al. 2007).
The quality factor can be attributed to the radial extent of the
transition zone where kHz QPOs produce (see Fig. \ref{model}).
In this letter, we investigate how this radial extent leads to the
profile for upper kHz QPO, which is described in Section 2, where
the comparison between theory and observations is also presented.
Section 3 contains the conclusions and discussions.

\section{Formation of High Q Factor and Its Variation with Frequency}

According to the MHD Alfv$\grave{e}$n wave oscillation model of kHz
QPO, the MHD turbulence by the shear flow in the accretion disk
(e.g., Ruediger \& Pipin 2000) will trigger the strong variation of
plasma energy density and ignite the shear Alfv$\acute{e}$n wave
motion along the orbit with the loop length of circumference $2 \pi
r$ at the Keplerian disk orbit radius $r$ (Zhang et al. 2007). In
the process of NS accretion, there is a transition from the
spherical accretion with high mass density to a polar cap accretion
with low mass density (Zhang 2004). It is assumed that this
transition occurred at a certain radius which is called preferred
radius, where a MHD tube loop may be formed to conduct the accreted
matter to the polar cap of star. This critical transition may give
rise to MHD turbulence and perturb the field lines, thus excite
Alfv$\grave{e}$n wave oscillation. In such a scenario, the lower and
upper frequency correspond to the AWOF with the spherical accretion
mass density which coincides with the Keplerian orbital frequency
$\nu_k$ and the AWOF with the polar accretion mass density,
respectively. At the preferred radius $r$ which is defined by the
magnetic pressure matching the ram pressure, these two frequencies
read (Zhang 2004), respectively,
\begin{eqnarray}
\label{upper}\nu_2 &=& \sqrt{\frac{GM}{4\pi^2r^3}} = 1850 (Hz) AX^{\frac{3}{2}},\\
\label{lower}\nu_1 &=& \nu_{A}(S_p) = \nu_2\sqrt{\frac{S_P}{S_r}} =
\nu_2 X \sqrt{1-\sqrt{1-X}}.
\end{eqnarray}
Here, the NS mass $M$ is in units of solar mass, and ${\rm \nu_{A}
\propto \sqrt{S}}$ is the AWOF where the area $S$ representing the
spherical area ${\rm S_r = 4\pi R^2}$ or the polar cap area ${\rm
S_P = 4\pi R^2(1-\sqrt{1-X})}$, respectively. ${\rm X=\frac{R}{r}}$
is the ratio between star radius $R$ and disk radius $r$. ${\rm A =
(m/{R_6^{\ 3}})^{1/2}}$ with ${\rm R_6 = R/10^6(cm)}$ and $m$ the
mass $M$ in the units of solar masses.

We ascribe the upper kHz QPO frequency profile to the X-ray flux
strong variations from transition layer, as shown in the upper panel
of Fig.(\ref{model}). Thus, the range of the upper frequency
($\delta \nu$) should be the indication of transition layer radial
extent ($\delta r$ or in terms of the scaled quantity $X$). \be
\delta \nt = \nt {3\delta X \over 2X} = \nt {3\delta r \over 2r}.
\ee If we set the QPO FWHM $w$ as corresponding to the scale of
transition layer width, i.e. ${\rm w \sim \delta \nu}$, then the
quality factor for upper frequency (hereafter upper Q-factor) can be
written as,
\begin{equation}
Q_2 = \frac{\nu_2}{\delta\nu_2} = \frac{2}{3}\frac{r}{\delta r}.
\end{equation}

We consider a transition layer (see upper panel of Fig.
(\ref{model})) between the innermost Keplerian orbit and the
magnetosphere (Elsner \& Lamb 1977; {\bf Naso \& Miller 2010ab}), in
which the transition for radial velocity of accretion flow from a
Keplerian to a corotation with the NS may occur (Titarchuk, Lapidus
\& Muslimov 1999 hereafter TLM; Titarchuk \& Osherovich 1999). But
in this region the formation of kinks and shocks due to the
inhomogeneous density of flow and supersonic motions of accretion
flow may be responsible for a super-Keplerian rotation (TLM). After
a self-adjustment which is the function of this layer, the coupling
of the sub-Keplerian flow with the super-Keplerian rotation still
leads to a Keplerian zone which contributes to the radial drift.
When the accreted matter fall into the innermost Keplerian orbit and
transition zone, these plasma may strike the magnetospheric boundary
and bend the field lines, which lead to the change of the
magnetospheric shape due to the strong ram pressure and can excited
the Alfv$\grave{e}$n wave oscillation. In the meantime, the plasma
is threaded by the field lines and fall onto the polar cap. However,
on the magnetospheric boundary, the accreted matter carries
different velocities and has different density from the former
arrivals, which bring into some instabilities, such as
Kelvin-Helmholtz instability, Rayleigh-Taylor instability, and the
inhomogeneities caused by the azimuthal component of the field
trapped inside the inner regions of the disk (Romanova, Kulkarni \&
Lovelace 2007). The Kelvin-Helmholtz instability seems to be less
significant (Rast$\ddot{a}$tter \& Schindler 1999), and we just
focus on the Rayleigh-Taylor instability (Kulkarni \& Romanova
2008). The unstable disc-magnetosphere boundary results in the
penetration of the magnetosphere by the disc matter in kinds of
forms. In addition, the disc matter treading the field lines into
the magnetosphere from polar cap may move to the equatorial region
and swell the magnetosphere. On the grounds of this idea and the
geometry of accretion disc boundary layer (Regev \& Hougerat 1988),
the radial extent direction can be written as (Elsner \& Lamb 1977;
{\bf Naso \& Miller 2010ab}),
\begin{equation}
\delta r = \alpha (r - R),
\end{equation}
where the quantity (r-R) stands for the scale of the
magnetosphere-disk boundary to stelar surface, and $\alpha < 1$ is a
constant ratio coefficient. Therefore, we can obtain,
\begin{equation}
Q_2 = \frac{2}{3\alpha}(1-\frac{R}{r})^{-1} =
\frac{2}{3\alpha}(1-X)^{-1}.
\end{equation}
Using the definition of X and Eq. (\ref{upper}), we can write Q$_2$
with respect to the upper frequency, {\bf \be Q_2 \sim
\alpha^{-1}[1-(\frac{\nu_2}{1850 A})^{\frac{2}{3}}]^{-1}. \ee} For
the detected twin kHz QPOs, the mass density parameter $A$ is found
to be about 0.7 (e.g. Sco X-1) (Zhang 2004; Zhang et al. 2007),
except for the two unusual X-ray millisecond pulsar cases SAX
J1804.5-3654 and XTE J1807-294, A=0.45 and 0.4 respectively (Zhang
et al. 2010). In most cases (except Cir X-1), the position parameter
X=R/r lies in the range from 0.7 to 0.92, or the kHz QPO emission
position radius is from r=1.1R to r=1.4R (Zhang et al. 2010).
Averagely, we have the following expression to evaluate the upper
Q-factor, {\bf \begin{equation} Q_2 \sim
\alpha^{-1}[1-(\frac{\nu_2}{1300 ({\rm Hz})})^{\frac{2}{3}}]^{-1}.
\label{q2}
\end{equation}}

\begin{figure}
\includegraphics[width=8cm]{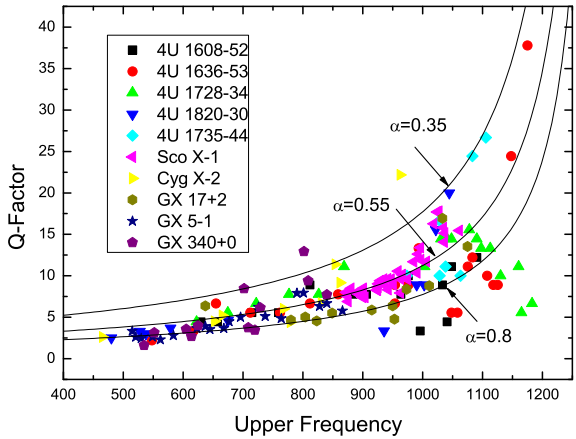}
\caption{Plot for the upper Q-factor versus its frequency. The data
are provided by D. Barret and M. Mendez, which have been exploited
and discussed in the references (Barret et al. 2005abc, 2006, 2007,
2008; Boutelier et al. 2009, 2010; Mendez 2006). The theoretical
curves of upper Q-factor $Q_{2}$ of Eq. (\ref{q2}) are plotted with
the parameter $\alpha = 0.35, 0.55, 0.8$ and the averaged mass
density parameter A=0.7 (Zhang 2004; Zhang et al. 2007).}\label{Q}
\end{figure}

\begin{figure}
\includegraphics[width=8cm]{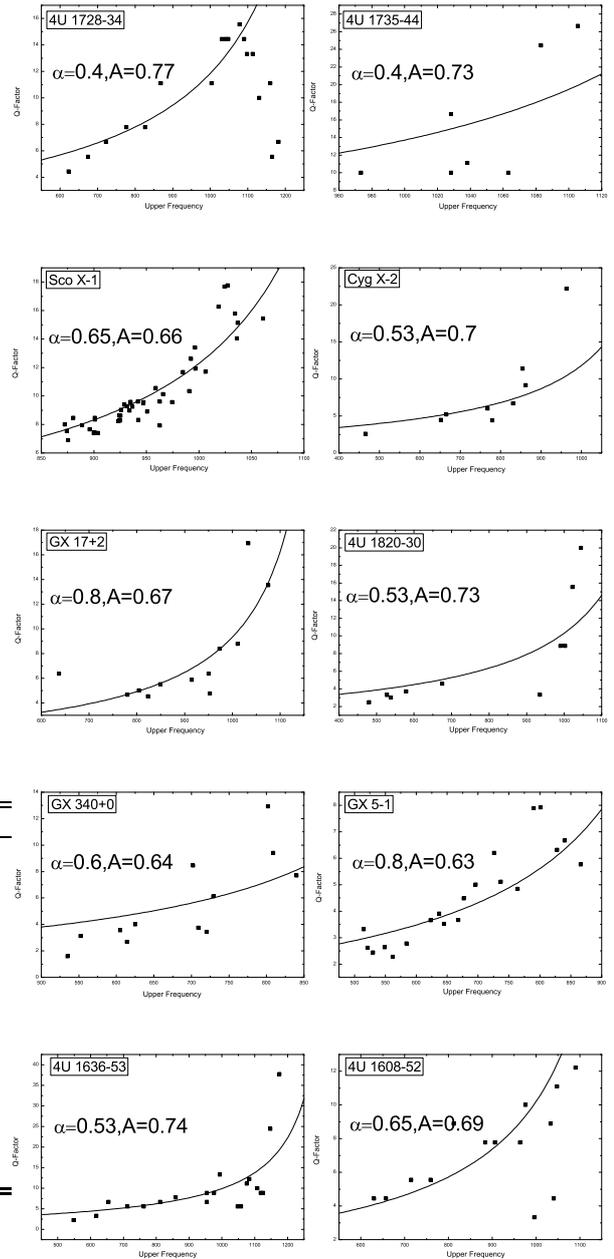}
\caption{The same meaning with  Fig. \ref{Q} but for separated
sources (five Atoll and five Z sources). The model curves are
plotted with the different extent parameter $\alpha$ and averaged
mass density parameter $A$ that were derived from the detecting data
(Zhang et al. 2007). }\label{sep-Q}
\end{figure}

To make a comparison between the model and the detections of the
upper Q-factor, we firstly put the sources whose spectrum of kHz
QPOs displayed the high coherence together in Fig. \ref{Q}. {\bf We
choose the parameter $\alpha = 0.35, 0.55, 0.8$, respectively, and
plot the theoretical curves. Then we fit the theoretical expression
for Q$_2$ (Eq. (\ref{q2})) to the observational data and find that
the model curve with $\alpha = 0.551$ fits the most data well, i.e.
the radial extent of the preferred radius is about half of the
thickness of the magnetosphere-disk boundary to stellar surface (see
Fig. \ref{Q}). It is can be seen that five Z sources and five Atoll
sources (Hasinger 1989, 1990) show different coherence trends, which
are described in Fig. \ref{sep-Q}, as the same meanings of Fig.
\ref{Q} but with the separate sources for a clearly presentation.
The fitting results are listed in Table 1.} For the Atoll source 4U
1728-34, there is an abrupt drop at $\nu \sim 1050$ Hz in the
Q-$\nu$ plot, then it is argued as an effect of innermost stable
orbit (Barret, Olive \& Miller 2005b,c; 2006) to arise this abrupt
kHz QPO profile changes.


\begin{table}
\centering
\begin{tabular}{c}
{Table 1. The fitting results for the observation data.}
\end{tabular}
\label{tab2}
\begin{tabular}{ccccc}
\hline\hline \ Source & \ \ \ $\alpha$ \ \ \ & \ \ \ error($\alpha$)
\ \ \ & \ \ \ $\chi^2/DoF$
 \ \ \ & error($\chi^2/DoF$) \\
\hline\\
Total & 0.551 & 0.030 & 3.729 & 0.601 \\
4U 1728-34 & 0.400 & 0.033 & 2.007 & 0.452 \\
4U 1735-44 & 0.400 & 0.027 & 14.095 & 5.337 \\
Sco X-1 & 0.650 & 0.007 & 1.391 & 0.539 \\
Cyg X-2 & 0.530 & 0.039 & 2.186 & 1.378 \\
GX 17+2 & 0.800 & 0.0340 & 6.015 & 1.463 \\
4U 1820-30 & 0.530 & 0.040 & 1.741 & 0.522 \\
GX 340+0 & 0.600 & 0.042 & 6.416 & 2.068 \\
GX 5-1 & 0.800 & 0.022 & 1.669 & 0.047 \\
4U 1636-53 & 0.530 & 0.027 & 4.540 & 0.450 \\
4U 1608-52 & 0.650 & 0.027 & 5.183 & 1.020 \\
\hline\hline
\end{tabular}
\end{table}


\section{Conclusions and Discussions}
\label{sec4}

We try to ascribe the profile of upper kHz QPO frequency to the
radial extent of its emission region at magnetosphere-disk
transition layer, and investigate the evolutions of its quality
factor as a function of upper frequency.
The central frequency of upper kHz QPO emits from a preferred radius
$r$, at which the magnetic pressure matches the ram pressure of the
inward disk material, and the radial extent of the preferred radius
contributes to the frequency extension of the upper kHz QPOs. The
narrower this extent, the higher coherence of QPO (or upper
Q-factor).
After making the comparisons between the model and detected Q-factor
data (Fig. \ref{Q}), we find that the radial extent $\delta r$
accounts for about half of the transition zone, extending from the
stellar surface to the magnetosphere-disk boundary, e.g. (r-R). This
means that X-ray flux from this radial extent primarily contributes
to the QPO peak profile (see illustration figure Fig.\ref{model}).
On the quantitatively discussion of the kHz QPO emission region, we
refer to the recent result by Zhang (2010) that the kHz QPOs of most
sources emit from the positions at r $\sim$ 20 km if stellar radius
R=15 km is assumed (excluding Cir X-1, its emission position is
slightly far from star, e.g. r $\sim 30$ km), thus the radial extent
of contributing to upper frequency is about 3 km.

While making comparisons of model to the Q-factors of five Atoll and
five Z sources separately, as shown in Fig. \ref{sep-Q}, we notice
that there exist a big discrepancy in 4U 1728-34 at high frequency,
more errors in Cyg X-2 and GX 340+0. On the unusual abrupt drop of
upper Q-factor of 4U 1728-34, it is ascribed as the effect of ISCO
(Barret, Olive \& Miller 2006). On the upper Q-factors of Cyg X-2
and GX 340 + 0, we argue that there might not only exists a radial
extent but also a vertical extent to contribute to the QPO emission
regions, and if the thickness extent of transition layer is
considered it might arise new ingredients on the QPO profiles.
On the thickness of accretion disk and QPO phenomenon, we refer to
the recent work by Chakrabarti et al. (2009). We think that the
complete description of QPO profile should take both radial and
thickness extents around preferred radius into account, since the
pileup blob of accreting plasma interacts with magnetosphere to
produce the fruitful X-rays in a spacial domain.
Moreover, it is remarked that our model on the profile of kHz QPO is
concentrated to the upper frequency $Q_2$, which cannot be
automatically applied to the lower frequency $Q_1$, and in fact the
observations of Q$_2$ and Q$_1$ are apparently different.
We ascribe the profile width of $\nu_2$ to the radial width of the
transition layer of magnetosphere-disk, which is scaled by $\delta$r
= r-R, but the profile width of $\nu_1$ should be not related to
this width. Therefore, there does not exist a relation between Q$_1$
and Q$_2$, although a relation between the upper and lower
frequencies can be clearly expressed. It is a subsequent work to
figure out the mechanism for the profile of lower kHz QPO $Q_1$.

\section{acknowledgements}

It is a pleasure to thank S.K. Chakrabarti for discussions, to M.
Mendez and D. Barret for providing the QPO data. This work has been
supported by the National Natural Science Foundation of China (NSFC
10773017) and the National Basic Research Program of China
(2009CB824800).

\end{document}